\documentclass[11pt]{article}
\usepackage[utf8]{inputenc}
\usepackage[T1]{fontenc}
\usepackage[a4paper,bindingoffset=0cm,head=15pt,body={17cm,25cm},top=2.5cm]{geometry}
\usepackage{amsmath,amssymb}
\usepackage{graphicx}
\usepackage{color}
\usepackage{cite}
\usepackage{ifthen}
\usepackage{titlesec}
\usepackage{fancyhdr}
\usepackage{booktabs}
\usepackage[english]{babel}
\usepackage{hyperref}
\usepackage{titlecaps}
\usepackage{xparse}

\titleformat{\section}{\normalfont\bfseries}{\thesection}{2ex}{}
\titlespacing*{\section}{0pt}{2.5ex}{1ex}
\titleformat{\subsection}{\normalfont\bfseries}{\thesubsection}{2ex}{}
\titlespacing*{\subsection}{0pt}{2.5ex}{1ex}
\newcommand{\E}{\ensuremath{\mathrm{e}}}

\newtheorem{definition}{Definition}

\newtheorem{theorem}[definition]{Theorem}

\newcounter{authorcounter}
\newcounter{institutecounter}

\newcommand{\wavesauthorlist}{}
\newcommand{\wavesaddresslist}{}
\newcommand{\wavesemail}{}
\newcommand{\wavesfootnotes}{}
\newcommand{\wavesauthorpre}{}

\def\theNumberTest#1{%
  \if\relax\detokenize\expandafter{\romannumeral-0#1}\relax
    true%
  \else
    false%
  \fi
}

\NewDocumentCommand{\wavesspeaker}{ O{} O{} m m}{%
    \ifthenelse{\value{authorcounter} > 1}{%
      \renewcommand{\wavesauthorpre}{, }%
    }{%
      \renewcommand{\wavesauthorpre}{}%
    }%
    \ifx\relax#1\relax
      \renewcommand{\wavesfootnotes}{}
    \else   
      \renewcommand{\wavesemail}{$^\ast$Email: #1}%
      \renewcommand{\wavesfootnotes}{, \ast}
    \fi
    \ifthenelse{\equal{\theNumberTest{#4}}{true}}{%
      \edef\wavesauthorlist{\wavesauthorlist%
        \wavesauthorpre{}\underline{#3}$^{#2%
        }$%
      }%
    }{%    
      \edef\wavesauthorlist{\wavesauthorlist%
        \wavesauthorpre\underline{#3}$^{\arabic{authorcounter}%
        }$%
      }
      \edef\wavesaddresslist{\wavesaddresslist% 
        \par%
        $^{\arabic{authorcounter}}$#4%
      }%      
      \stepcounter{authorcounter}%
      \stepcounter{institutecounter}
    }%

    \ifx\relax#2\relax
          \edef\wavesauthorlist{\wavesauthorlist%
        \wavesauthorpre$^{\wavesfootnotes%
        }$%
      }
    \else
    \ifthenelse{\equal{\theNumberTest{#2}}{true}}{%
      \edef\wavesauthorlist{\wavesauthorlist%
        \wavesauthorpre{}$^{,#2\wavesfootnotes%
        }$%
      }%
    }{%    
      \edef\wavesauthorlist{\wavesauthorlist%
        \wavesauthorpre$^{,\arabic{institutecounter}\wavesfootnotes%
        }$%
      }
      \edef\wavesaddresslist{\wavesaddresslist% 
        \par%
        $^{\arabic{institutecounter}}$#2%
      }%      
      \stepcounter{institutecounter}%
    }%
    \fi
  \ignorespaces
}

\NewDocumentCommand{\wavesauthor}{ O{} O{} m m}{%
    \ifthenelse{\value{authorcounter} > 1}{%
      \renewcommand{\wavesauthorpre}{, }%
    }{%
      \renewcommand{\wavesauthorpre}{}%
    }%
    \ifx\relax#1\relax
      \renewcommand{\wavesfootnotes}{}
    \else   
      \renewcommand{\wavesemail}{$^\ast$Email: #1}%
      \renewcommand{\wavesfootnotes}{, \ast}
    \fi%
    \ifthenelse{\equal{\theNumberTest{#4}}{true}}{%
      \edef\wavesauthorlist{\wavesauthorlist%
        \wavesauthorpre{}#3$^{#4%
        }$%
      }%
    }{% 
      \edef\wavesauthorlist{\wavesauthorlist%
         \wavesauthorpre{}#3$^{\arabic{institutecounter}%
         }$%
      }
      \edef\wavesaddresslist{\wavesaddresslist%
        \par%
        $^{\arabic{institutecounter}}$#4%
      }%
      \stepcounter{authorcounter}
      \stepcounter{institutecounter}%
    }%

    \ifx\relax#2\relax
          \edef\wavesauthorlist{\wavesauthorlist%
        $^{\wavesfootnotes%
        }$%
      }
    \else
    \ifthenelse{\equal{\theNumberTest{#2}}{true}}{%
      \edef\wavesauthorlist{\wavesauthorlist%
        $^{,#4\wavesfootnotes%
        }$%
      }%
    }{%    
      \edef\wavesauthorlist{\wavesauthorlist%
        $^{,\arabic{institutecounter}\wavesfootnotes%
        }$%
      }
      \edef\wavesaddresslist{\wavesaddresslist% 
        \par%
        $^{\arabic{institutecounter}}$#2%
      }%      
      \stepcounter{institutecounter}%
    }%
    \fi
  \ignorespaces
}

\fancypagestyle{plain}{%
  \fancyhead{}
}

\newenvironment{wavespaper}[3]{%
  \renewcommand{\wavesauthorlist}{}%
  \renewcommand{\wavesemail}{}%
  \setcounter{authorcounter}{1}%
  \setcounter{institutecounter}{1}%
     #2
  \twocolumn[
    \begin{center}
     \bfseries
     #1
     \bigskip

     \wavesauthorlist
     \mdseries
     \smallskip

     \wavesaddresslist
     \smallskip
 
     \wavesemail
    \end{center}%
  ]
}{%
}

\def\R{\mathbb{R}}

\def\uv{\mathbf{u}}
\def\rv{\mathbf{r}}
\def\vv{\mathbf{v}}

\def\uvn{{\mathbf{u}_0}}

\def\Orefv{\mathbf{\Omega}_\text{ref}}
\def\Oref{\Omega_\text{ref}}

\def\E{\mathbb{E}}
\def\cov{\mathrm{Cov}}

\begin{document}

\begin{wavespaper}{%
% Put the paper title here e.g
Inferring solar differential rotation and viscosity via passive imaging with inertial waves
}{%
% Put the list of authors here, using \wavesauthor or \wavesspeaker commands
  \wavesspeaker[nguyen@mps.mpg.de]{Tram T. N. Nguyen}{Max-Planck Institute for Solar System research, G\"ottingen, Germany}
    \wavesauthor[][Institute for Numerical and Applied Mathematics, University of G\"ottingen, Germany]{Thorsten Hohage}{1}
  \wavesauthor[]{Damien Fournier}{1}
  %\wavesauthor[][Center for Space Science, NYUAD Institute, New York University Abu Dhabi]{Laurent Gizon}{Institute for Astrophysics, University of G\"ottingen, Germany, \textcolor{red}{add 1}}
  \wavesauthor[][Institute for Astrophysics, University of G\"ottingen, Germany]{Laurent Gizon}{1}
   
}

\section*{Abstract}
The recent discovery of inertial waves on the surface of the Sun offers new possibilities to learn about the solar interior. These waves are long-lived with a period on the order of the Sun rotation period ($\sim$27 days) and are sensitive to parameters deep inside the Sun. They are excited by turbulent convection, leading to a passive imaging problem. In this work, we present the forward and inverse problem of reconstructing viscosity and differential rotation on the Sun from cross-covariance observations of these inertial waves.
\smallskip

\noindent\textbf{Keywords:} inverse problems, passive imaging, inertial waves, correlation data.

\section{Introduction}
Helioseismology aims at recovering parameters in the solar interior from surface observations using mostly acoustic \emph{modes} (eigenfunctions). These modes have maximum sensitivity close to the surface and inferring deep inside the Sun is an extremely difficult task. The recent discovery of many inertial modes  \cite{GizonInertial21} opens up new opportunities for helioseismology. A linearized analysis of purely toroidal (i.e.~divergence-free) modes on the surface of the Sun can already explain several observed features, such as the eigenfrequencies and eigenfunctions of %observed
 Rossby modes and %the presence 
 of high-latitude modes \cite{Damien22}. We use this forward model with an additional source %term 
 to take into account the stochastic excitation of the waves, and solve an inverse problem to recover the solar differential rotation and the viscosity.

As this source is stochastic% and vanishes in expectation
, we approach this from a \emph{passive imaging} perspective; that is, we replace the source by its auto-correlation, which is an accessible entity known as the \emph{source strength}, %and similarly also auto-correlate the observed data, 
leading to a higher-dimensional problem and exacerbated nonlinearity, for which special strategies are required.

\section{Solar inertial waves}
We consider the dynamic of a moving particle in an incompressible fluid. Its velocity $\vv=\vv(\rv,t)$ is described by the Navier-Stokes equation
\begin{align*}%\label{eq-motion}
\begin{split}
&\rho\left(\frac{\partial \vv}{\partial  t}+ \vv\cdot\nabla\vv \right)\\&\qquad=-\nabla p +\nabla\cdot(\mathbf{\rho\gamma\boldsymbol\tau})+\mathbf{f}- 2\Orefv\times\vv.
\end{split}
\end{align*}
The restoring forces on the right hand side include acoustic pressure $p$, viscous stress $\mathbf{\boldsymbol\tau}:=\nabla\vv+(\nabla \vv)^T$ and external source $\mathbf{f}$ (e.g. random source); $\rho$ denotes density. The Coriolis force  $(2\Orefv\times\vv)$ results from observing the Sun in a rotating reference frame, often chosen as the Carrington frame with the angular velocity $\Oref=14.7^0$/days. 

We decompose $\vv$ into a background flow %(differential rotation at depth $r$) 
$\uvn$ and a perturbation (wave field) $\uv$ in spherical coordinates
\[\vv=\uvn+\uv,\quad \uvn:=(\Omega(r,\theta)-\Oref)[0;0;r\sin\theta],\]  
and suppose that the fluid is strongly stratified \cite{Watson81}. Considering the horizontal motion and expressing it via the stream function $\Psi$ yields \[\rho\uv=\nabla\times \mathbf{\Psi}, \quad \mathbf{\Psi}:=[\Psi(t,\theta,\phi);0;0].\] Taking the Fourier transform in time and longitude, we arrive at the scalar equation for each frequency-longitude wave number, i.e.~each ($\omega,m$), as%, the inertial oscillation satisfies the scalar equation on $I:=(0,\pi)$
\begin{equation}\label{eq-model}
\begin{split}
&-\gamma \Delta_m^2\Psi-i\omega\Delta_m\Psi+im\beta_\Omega\Delta_m\Psi -im\alpha_\Omega\Psi=f \\[0.5ex]
&\text{with}\,\,\,\gamma\in\R, \quad\beta_\Omega(\theta):=\Omega(\theta)-\Oref\quad \theta\in (0,\pi)\\
&\alpha_\Omega(\theta):=\frac{1}{r^2\sin\theta}\frac{d}{d\theta}\left(\frac{1}{\sin\theta}\frac{d}{d\theta}(\Omega(\theta)\sin^2\theta)\right),\\
&\Delta_m\Psi:=\frac{1}{r^2\sin\theta}\frac{d}{d\theta}\left(\sin\theta\frac {d }{d \theta }\Psi\right)-\frac{m^2}{r^2\sin^2\theta}\Psi,
\end{split}
\end{equation}
where $\Delta_m$ is the Laplace-Beltrami operator on the sphere. %, and $\gamma$ is assumed to be constant on the Sun's surface. 
Supposing that the %velocity 
flow is continuous at the poles, we endow %\eqref{eq-model} with the Cauchy 
at the boundaries
\begin{equation}\label{eq-boundary}
\Psi(0)=\Psi(\pi)=0,\quad \Psi'(0)=\Psi'(\pi)=0.
\end{equation}
The system \eqref{eq-model}-\eqref{eq-boundary} is the underlying model for the inverse problem of recovering viscosity $\gamma$ and differential rotation $\Omega$. Equation \eqref{eq-model} reduces to that studied in \cite{Watson81} if $\gamma=0$ and in \cite{Damien22} if $f=0.$

\section{Correlation-based passive imaging}
On the Sun, the source of excitation $f$ is passive, i.e.~ambient noise near the surface of the convection zone. Consequently, the oscillation $\Psi$ is a realization of a random wave. If $f$ is spatially uncorrelated %at each frequency 
with $\E[f]=0$, then its cross-correlation between two locations %on the surface 
is
\begin{align*}%\label{cov_f}
\mathbb{C}\text{ov}[f,f](\theta,\theta')=\E[f(\theta)\overline{f(\theta')}]=:\Pi_f(\theta)\delta(\theta-\theta'),
\end{align*}
where $\overline{(\cdot)}$ denotes the complex conjugate, $\delta$ is the Dirac function, and $\Pi_f$ represents the source strength. In this spirit, averaging $J$ correlated wave fields acquired separately yields
\begin{align*}%\label{cov_psi}
\mathbb{C}\text{ov}[\Psi,\Psi]\approx
\cov(\Psi)(\theta,\theta'):=\frac{1}{J}\sum_{j=1}^J \Psi_j(\theta)\overline{\Psi_j(\theta')}, 
\end{align*}
referred to as the \emph{empirical reprocessed data}. It is important to note that the source term $f$ is typically not available, with only the source strength $\Pi_f$ assumed known.  

We formulate the passive imaging problem for solar viscosity and differential rotation as
\begin{align*}
\text{Find }(\gamma,\Omega)\in\R\times H^2(0,\pi)&:\quad \text{Cov}(\Psi)=y^\mathrm{obs}\quad\\ &\text{s.t. } \Psi \text{ solves \eqref{eq-model}-\eqref{eq-boundary}}
\end{align*}
given %the source strength 
$\Pi_f$ and noisy observed data $y^\mathrm{obs}$.

\section{Inversion numerical results}

%Based on Theorem \ref{theo}, 
We implemented an accelerated Nesterov Landweber algorithm to simultaneously reconstruct the scalar viscosity $\gamma$ and latitude-dependent rotation $\beta=\Omega$ (i.e.~$\Oref=0$) from a single frequency-longitude wave number.%, i.e.~the model \eqref{eq-model}-\eqref{eq-boundary} at a  single $(\omega,m)$.

%The algorithm in our setting reads as
%\begin{equation*}
%\begin{split}
%&x_{-1}=x_0:=(\gamma_\mathrm{int},\beta_\mathrm{int}):=(10\gamma_\mathrm{true},0)\\
%&z_k = x_k + \frac{k-1}{k+\alpha-1}\left(x_k-x_{k-1}\right) \qquad \alpha=3\\
%& x_{k+1} = z_k-\mu_k F'(z_k)^*\left( F(z_k)-y^\mathrm{obs} \right) \quad k\leq K
%\end{split}
%\end{equation*}
%where $F'(z_k)^*$ is as in Theorem \ref{theo}, iii), and the step size $\mu_k$ is determined via backtracking line search.
Figure \ref{fig-image} shows the reconstruction results (top) and the corresponding covariance images (bottom). We initiate the algorithm with $\beta_\text{init}=0$, i.e.~no prior knowledge, and $\gamma_\text{init}$ very far from the ground truth. Despite this, the reconstruction approximates the ground truth extremely well. The algorithm required 200 iterations over 2 seconds on an i7-1255U CPU (4.70 GHz).

\begin{figure}[htb]\centering
\includegraphics[scale=0.5]{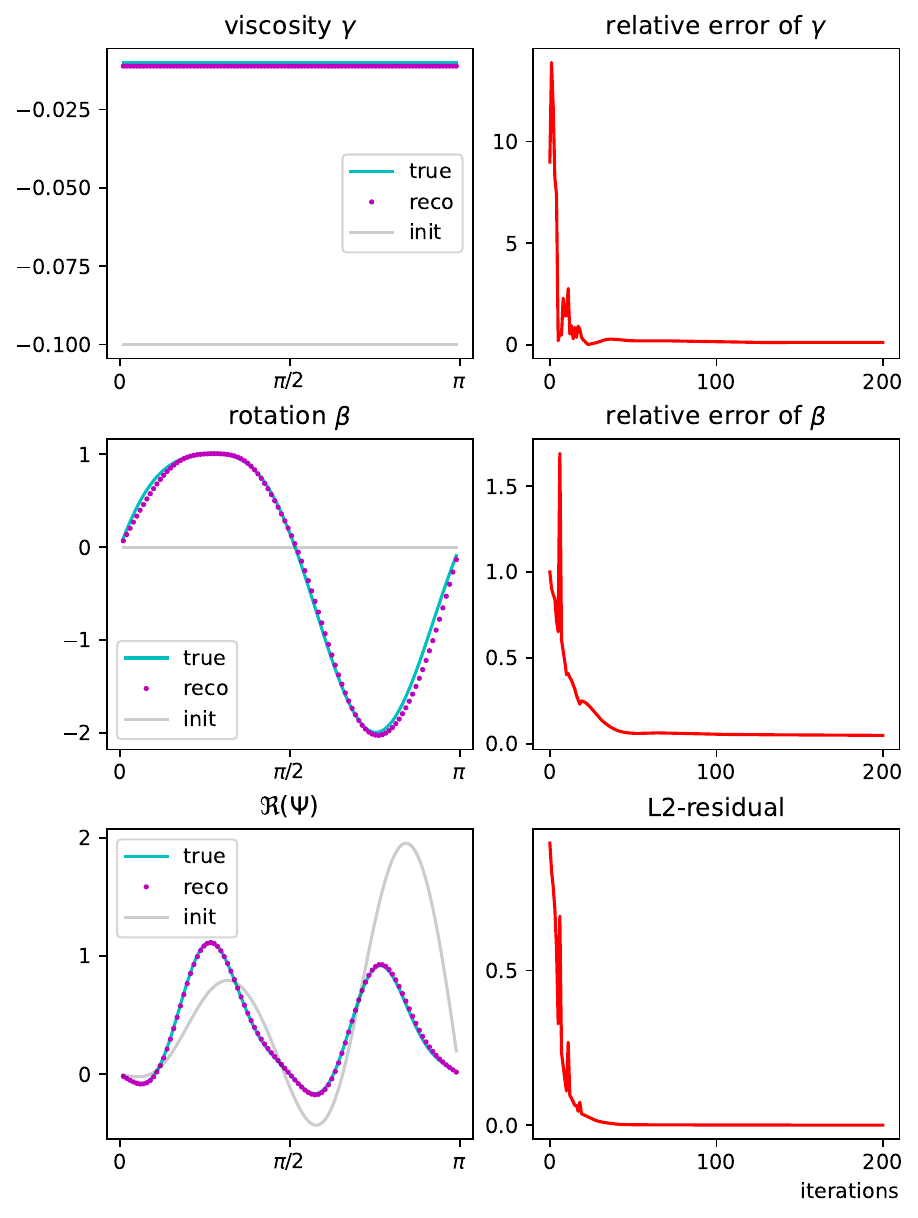}\\
\hspace{-1cm}
\includegraphics[scale=0.29]{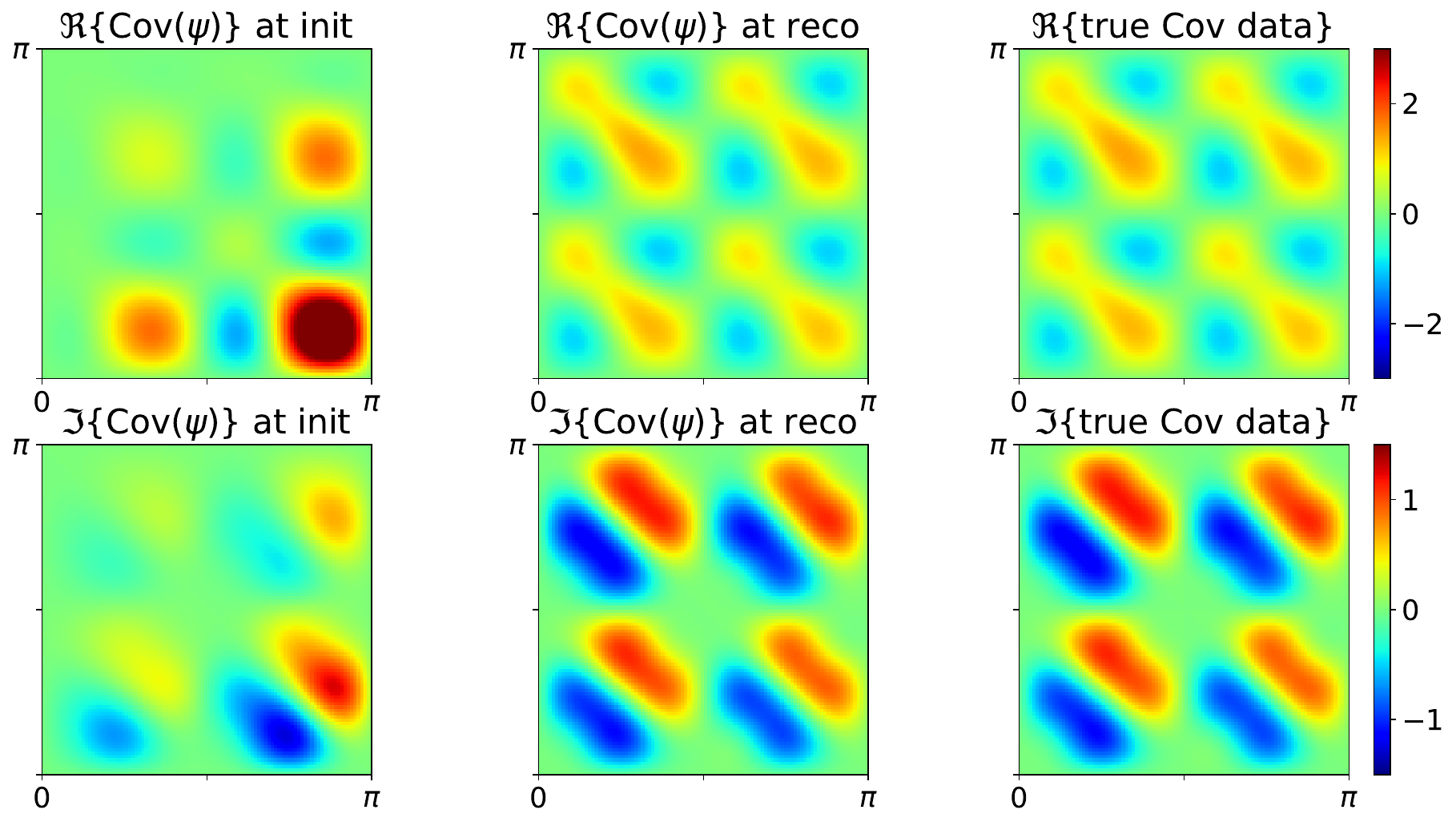}
\caption{Simultaneous reconstruction of viscosity/rotation (top) and resulting covariance images (bottom). $\text{Error}_x=\|x_\text{reco}-x_\text{true}\|/\|x_\text{true}\|$.}\label{fig-image}
\end{figure}

\paragraph{Outlook.} Going forward, we will perform imaging with solar-like parameters and real data acquired from the
Helioseismic and Magnetic Imager on board the Solar Dynamics Observatory.

 %------------------------------------------------------------------------
% --------------- DO NOT EDIT THE FILE BELOW THIS LINE -------------------
% ------------------------------------------------------------------------

\end{wavespaper}

\end{document}